\documentclass[conference,compsoc]{IEEEtran}
\IEEEoverridecommandlockouts
\ifCLASSOPTIONcompsoc
  \usepackage[nocompress]{cite}
\else
  \usepackage{cite}
\fi
\ifCLASSINFOpdf
\else
\fi
\usepackage{epsfig,endnotes, amsmath, amsfonts, amsthm, svg, bbm}
\usepackage{graphicx}
\begin{document}
\newtheorem{theorem}{Theorem}[section]
\newtheorem{definition}{Definiton}[section]
\newtheorem{corollary}{Corollary}[section]
\newcommand{\MyMod}[1]{\ (\mathrm{mod}\ #1)}
\title{Differentially-private Continual Releases against Dynamic Databases}

\author{\IEEEauthorblockN{Mingen Pan}
\IEEEauthorblockA{Independent Researcher \IEEEauthorrefmark{1} \thanks{\IEEEauthorrefmark{1} Currently working at Google. This work was conducted independently and does not reflect the views or endorsement of Google.}\\
Email: mepan94@gmail.com}
}


% compsoc conferences compared to traditional conferences), use this
%Telephone: (800) 555--1212, Fax: (888) 555--1212}


\maketitle

\begin{abstract}
Prior research primarily examined differentially-private continual releases against data streams, where entries were immutable after insertion. However, most data is dynamic and housed in databases. Addressing this literature gap, this article presents a methodology for achieving differential privacy for continual releases in dynamic databases, where entries can be inserted, modified, and deleted. A dynamic database is represented as a changelog, allowing the application of differential privacy techniques for data streams to dynamic databases. To ensure differential privacy in continual releases, this article demonstrates the necessity of constraints on mutations in dynamic databases and proposes two common constraints. Additionally, it explores the differential privacy of two fundamental types of continual releases: Disjoint Continual Releases (DCR) and Sliding-window Continual Releases (SWCR). The article also highlights how DCR and SWCR can benefit from a hierarchical algorithm, initially developed by \cite{dwork2010continual, chan2011private}, for better privacy budget utilization. Furthermore, it reveals that the changelog representation can be extended to dynamic entries, achieving local differential privacy for continual releases. Lastly, the article introduces a novel approach to implement continual release of randomized responses.
\end{abstract}


% \fi
\IEEEpeerreviewmaketitle

\section{Introduction}

We continually ask questions like "What is the current unemployment rate in the US?" or "How many people are currently infected with COVID-19?" because the subjects of these questions (i.e. people) are constantly changing. Answers from just a short while ago may no longer accurately reflect the current state of the subjects. It's important to notice that answering the questions mentioned above requires personal data. Continuously releasing data derived from personal information could result in a significant loss of privacy for individuals. For instance, \cite{dinur2003revealing} showed that a database could be reconstructed given enough perturbed aggregations of individual data. Additionally, \cite{molina2010meter} demonstrated that the continual release of a resident's smart electric meter data resulted in a breach of their privacy.

Differential privacy \cite{dwork2006privacy, dwork2014privacy} has emerged as the leading standard for protecting individual privacy in data analysis. Differential privacy requires a query to have similar output distributions for two adjacent inputs, and formally defines an inequality to measure the similarity. Therefore, the participation of an entry in a query will have little impact on the output of the query, as if the entry is not even included, which protects the privacy of the entry.

Previous studies have integrated differential privacy with continual release. PINQ\cite{mcsherry2009PINQ} first introduced the parallel composition of differential privacy, which proved that the queries against disjoint subsets of a database would have a total privacy loss equivalent to the max privacy loss among the queries. Given the parallel composition, the differential privacy of a continual release can be bounded if each query of the continual release is against one of the disjoint subsets of a data stream. However, the sum of the results of a continual release will have the variance accumulated linearly. To slow down the accumulation of variance, \cite{dwork2010continual} and \cite{chan2011private} independently propose differentially-private hierarchical algorithms to continually count a binary data stream while accumulating the variance logarithmically through time. \cite{qardaji2013} analyzed the variance of the hierarchical algorithm and proposed an optimized branching factor to best utilize the privacy budget. \cite{perrier2018PAK, wang2021continuous} introduced a few fine tunings into the hierarchical algorithm, such as truncating data, to better utilize the privacy budget for continual releases. \cite{zhang2023differentially} extended the hierarchical algorithm to ensure user-level differential privacy. In their method, each user can only contribute a constant number of records to the data stream, which is similar to the at-most-$k$-mutation constraint introduced later in the paper.

All the researches above only study the continual release on a data stream, where entries are immutable once they are inserted. However, dynamic data can also be stored in a database. There is other research focused on dynamic databases. \cite{cummings2018growing} studied the continual release against a growing database and found that the accuracy of some algorithms increases as the database grows. They proposed continually querying a growing database with exponentially decreasing privacy budgets just enough to achieve a required accuracy, which leads to a convergence of the cumulative privacy loss. However, this method requires an exponentially growing database. Also they do not leverage the sequential composition and treat the database at different times as a new database. Another research on differential privacy of dynamic databases \cite{lecuyer2019privacy, luo2021privacy} has proposed an idea to partition a database into blocks, each with its own privacy budget. Incoming data is written into a new block and old data could be deleted from the old blocks. However, their research did not consider the mutation of an entry. Also, they focused on privacy budget management rather than the algorithms of continual release.

Taking into account the limitations of prior research, this paper proposes an innovative methodology for achieving differential privacy in continual releases against dynamic databases. We represent a dynamic database as a changelog, which is a stream of immutable mutation records, so the prior research on the differential privacy of data streams can be applied to dynamic databases. However, the differential privacy on a changelog is different from that on a data stream, because one entry may have multiple mutations, leading to two adjacent databases having changelogs different in multiple places. To address this difference, this article derives the definition of differential privacy in terms of two adjacent changelogs and applies it to continual releases. The paper also recognizes that constraints on mutations are essential for enforcing differential privacy on a continual release. Without such constraints, the mutations of a single entry could affect every query of the continual release, leading to an unbounded privacy loss. To address this issue, this paper introduces two constraint types - “at-most $k$” and “time-bounded” - to limit the number of queries an entry can affect.

Then this paper examines two types of continual release: Disjoint Continual Release (DCR), whose queries receive disjoint sets of a changelog as input, and Sliding-Window Continual Release (SWCR), whose query filter shifts like a sliding window. For both types, the article examines their differential privacy against a dynamic database with the aforementioned constraints on mutations. In addition, this paper applies the hierarchical algorithm to the DCR to develop Hierarchical Disjoint Continual Release (HDCR). Like the traditional hierarchical algorithm \cite{dwork2010continual, chan2011private}, the variance of the aggregated results from a HDCR increases logarithmically through time when running linear queries. The article also demonstrates how to derive SWCR from HDCR and discusses how, with the same privacy loss, deriving SWCR from HDCR can yield better utility in certain conditions.

Apart from global differential privacy, this paper also studies the local differential privacy \cite{cormode2018LDP} of continual releases against individual dynamic entries. \cite{rappor} was the first research to tackle this problem by memorizing queries and recomputing the queries only when the entries are changed, but \cite{ding2017collecting} pointed out that this method would reveal the mutation time of entries. Such information could have serious implications, such as exposing a patient's medical history. To address this problem, \cite{joseph2018local} proposed a differentially-private local voting mechanism that safeguards the mutation time of entries. This paper presents a new methodology: similar to a database, the lifetime of an entry can be represented by a changelog too; this paper provides a proof that a continual release against a dynamic entry can have its privacy loss bounded if the entry has constraints on mutations. Additionally, the paper extends the global differential privacy guarantees provided by DCR, HDCR, and SWCR to the setting of local differential privacy.

Also, this paper builds on the privacy guarantee of DCR to propose a method for continually releasing randomized responses while preserving local differential privacy for dynamic entries. Randomized response \cite{warner1965randomized} is a technique to map the true answer from an entry to a perturbed value. The randomized responses from a group of entries could derive the distribution of the true answer while preserving the privacy of individual inputs. This technique has been widely utilized in various fields, including medical surveys \cite{chow1972RR, goodstadt1975RR}, behavioral science \cite{donovan2003RR}, and biological science \cite{st2012RR}. As the true answers from dynamic entries can change over time, it is crucial to continuously conduct randomized-response surveys to monitor their distribution. This paper proposes an approach to continually conduct randomized-response surveys on the mutations of the true answers, which unbiasedly estimate the histogram of the true answers through the whole continual release. Moreover, by adapting HDCR, the variance of the estimated histograms grows only logarithmically with time.

\section{Background}
\subsection{Notion}

The definition of a symbol (e.g. $x$) will be extended to later sections until the same symbol is redefined. Bold symbols (e.g. $\mathbf{x}$) denote a sequence, where the $i$-th element is either denoted by a subscript (e.g. $x_i$) or a bracket (e.g. $\mathbf{x}[i]$). $[n]$ denotes an integer sequence from zero to $n$, while $[n]^+$ denotes an integer sequence from 1 to $n$. $(x_i)_{i \in [n]^{+}}$ denotes a sequence $(x_1, x_2, ..., x_n)$. $\{x \in \mathbb{X} | f(x), g(x), ... \}$ denotes a set of elements in $\mathbb{X}$ constrained by $f(\cdot)$, $g(\cdot)$, and etc.

\subsection{Entry}
A database is a collection of entries from a data universe $\mathcal{X}$. The entries of a static database are constant, while the entries of a dynamic database may change over time. Formally, an entry $x$ is a function of time $t$. $x_t$ denotes the state of the entry $x$ at $t$, i.e. $x(t)$. If $t$ is outside the lifetime of $x$, we set $x(t) = \text{null}$. 

\subsection{Dynamic Database}

This article adapts the histogram representation of a database, $\bf{x} \in \mathbb{N}^{|\mathcal{X}|}$, where $\mathbf{x}[i]$ represents the number of the $i$-th entry of the data universe $\mathcal{X}$ in the database.

A dynamic database is a database that is subject to change. Each entry in the database has a lifecycle that includes insertion, potential modifications, and potential removal. We say that the database $\mathbf{x}$ contains an entry $x$, i.e. $x \in \mathbf{x}$, if $x$ will exist in the database at some point in time. We denote the snapshot of the database at time $t$ as $\mathbf{x}_t$. 

\subsection{Adjacent Databases}

Two databases are considered to be adjacent if they only differ by one entry. Formally, $\mathbf{x}_1, \mathbf{x}_2 \in N^{|\mathcal{X}|}$ are adjacent if and only if their hamming distance $|\mathbf{x}_1 - \mathbf{x}_2| \leq 1$.

\subsection{Differential Privacy (DP)}

A query $q \in \mathbb{Q}: \mathbb{N}^{|\mathcal{X}|} \rightarrow \mathbb{R}$ takes a database as input and outputs a randomized result in the range $\mathbb{R}$. For any two adjacent databases $\mathbf{x}$ and $\mathbf{x}'$, the query $q$ is $(\epsilon, \delta)$-DP if and only if for any $S \subseteq \mathbb{R}$, we have

\begin{equation}
    Pr[q(\mathbf{x}) \in S] \leq e^{\epsilon} Pr[q(\mathbf{x}') \in S] + \delta.
    \label{eq:dp_eq}
\end{equation}

\subsection{Sequential Composition}
\label{sec:sequential_composition}

When a database is queried multiple times with privacy loss of $(\epsilon_1, \delta_1), (\epsilon_2, \delta_2), ..., (\epsilon_n, \delta_n)$, the total privacy loss of the database can be represented by $SC_{i \in [n]^{+}} (\epsilon_i, \delta_i) $. Here, $SC$ represents a generic sequential composition algorithm, which could be the naive or advanced composition \cite{dwork2014privacy, kairouz2015composition}, or others. By default, the compositions used in this paper later are non-adaptive, so non-adaptive composition can also be applied here, e.g. \cite{sommer2018privacy, dong2020optimal}. This paper will also use $SC[(\epsilon_1, \delta_1), (\epsilon_2, \delta_2), ...]$ to denote the composition with the given privacy loss.

This paper will frequently apply sequential composition to multiple queries with the same privacy loss. Let $k \text{-} (\epsilon, \delta)$ denote $SC_{i \in [k]^{+}} (\epsilon, \delta)$, which represents $k$-fold sequential composition of $(\epsilon, \delta)$. This paper will also apply composition to a set of composite queries:

\begin{theorem}
    (Composition of Compositions)

    \begin{equation}
    \underset{i}{SC} \left( \underset{j \in f(i)}{SC} (\epsilon_{ij}, \delta_{ij}) \right) = \underset{i, j \in f(i)}{SC} (\epsilon_{ij}, \delta_{ij}),
\end{equation}

\noindent where $f(i)$ is a sequence of $j$ given $i$.
\end{theorem}

\noindent Proof: the theorem is self-explained; if we know all the base queries of the nested compositions, we could consider them holistically and apply the composition to them as a batch. 

Applying this theorem to the $k$-fold composition, and we get

\begin{corollary}
    \begin{equation}
        \underset{i}{SC} \; k_i \text{-} (\epsilon, \delta) = (\sum_i k_i) \text{-} (\epsilon, \delta) .
    \end{equation}
    \label{cor:nested_k_fold}
\end{corollary}

\subsection{Parallel Composition}

This section will introduce a more general form of the parallel composition \cite{mcsherry2009PINQ}. Suppose a sequence of queries, $\mathbf{q}$, each query $q_i$ of which has a filter $d_i$ and is $(\epsilon_i, \delta_i)$-DP. $x \in d_i$ denotes that $d_i$ accepts an entry $x$, and $x$ affects the output of $q_i$, while $x \notin d_i$ is the opposite. Also each query does not depend on the previous queries, i.e. non-adaptive.

Given an arbitrary database $\mathbf{x}$ and an entry $x$, the output probability of $\mathbf{q}$ is

\begin{equation}
    Pr(\mathbf{q}(\mathbf{x} + x)) = \prod_{i} Pr(q_i(d_i(\mathbf{x} + x)))
\end{equation}

\begin{equation}
    =  \prod_{i : x \in d_i}  Pr[q_i(d_i(\mathbf{x} + x))] \times \prod_{i : x \notin d_i}  Pr[q_i(d_i(\mathbf{x}))] .
\label{eq:parallel_composition_step_1}
\end{equation}

\noindent Define 

\begin{equation}
    L(q, x) = (\epsilon_{q, x}, \delta_{q, x}) = \underset{x \in d_i}{SC}  (\epsilon_i, \delta_i) ,
    \label{eq:L_q_x}
\end{equation}

\noindent which is the sequential composition of all the queries whose filters accept $x$. Then Eq. \eqref{eq:parallel_composition_step_1} can be written as

\begin{equation}
    \eqref{eq:parallel_composition_step_1}
    \le  \{ e^{\epsilon_{q, x}} \prod_{i : x \in d_i}  Pr[q_i(d_i(\mathbf{x}))] + \delta_{q, x} \} \times \prod_{i : x \notin d_i}  Pr[q_i(d_i(\mathbf{x}))].
\label{eq:parallel_composition_step_2}
\end{equation}

\noindent Given $\prod_{i : x \notin d_i}  Pr[q_i(d_i(\mathbf{x}))] \le 1$, we combine the first and second terms of the Eq. \eqref{eq:parallel_composition_step_2} and have

\begin{equation}
    \eqref{eq:parallel_composition_step_2}
    \le  e^{\epsilon_{q, x}} \prod_{i}  Pr[q_i(d_i(\mathbf{x}))] + \delta_{q, x} ,
    \label{eq:parallel_composition_step_3}
\end{equation}

\noindent Eq. \eqref{eq:parallel_composition_step_3} is a weaker form of differential privacy because $(\epsilon_{q, x}, \delta_{q, x})$ depends on $x$. If there exists a privacy loss $(\epsilon, \delta)$ independent of $x$ while satisfying Eq. \eqref{eq:parallel_composition_step_3}, then Eq. \eqref{eq:parallel_composition_step_3} is equivalent to the definition of differential privacy, and  $q$ is $(\epsilon, \delta)$-DP.

\begin{definition}
    $(\epsilon_1, \delta_1) \preceq (\epsilon_2, \delta_2)$ iff $\epsilon_1 \le \epsilon_2$ and $\delta_1 \le \delta_2$.
\end{definition}

\begin{theorem}
    $q$ is $[\sup_{x \in \mathcal{X}} L(q, x)]$-DP.
    \label{thm:parallel_composition}
\end{theorem}

\noindent Proof: define  $(\epsilon, \delta) = \sup_{x \in \mathcal{X}} L(q, x)$. For any $x \in \mathcal{X}$, $(\epsilon_{q, x}, \delta_{q,x}) \preceq (\epsilon, \delta)$, so

\begin{equation}
    \eqref{eq:parallel_composition_step_3} \le e^{\epsilon} \prod_{i}  Pr[q_i(d_i(\mathbf{x}))] + \delta. \qed
    \label{eq:parallel_composition_step_4}
\end{equation}

Though the parallel composition is proved with the $(\epsilon, \delta)$ definition, it also holds for other forms of DP such as Rényi DP \cite{mironov2017renyi}. The theorems derived from the parallel composition in the following sections are applicable to any forms of DP satisfying the parallel composition.

\subsection{Continual Release}

A continual release comprises a stream of queries executed sequentially. Let $\mathbf{q} = (q_i)_{i \in [n]^{+}}$ denote a continual release, where $q_i$ is the $i$-th query. $n$ denotes the size of a continual release, which can be any positive integers, or infinity if the continual release is unbounded.

A continual release is either adaptive or non-adaptive. The queries of an adaptive continual release depend on the results of the previous queries, while the queries of a non-adaptive continual release do not. For the purposes of this article, we will assume that the continual releases are non-adaptive by default.

\subsection{Linear Query}
\label{sec:linear_query}

Linear queries generally take the form of 

\begin{equation}
    E[q(\mathbf{x})] = \sum_{x \in \mathbf{x}} f(x).
    \label{eq:linear_query}
\end{equation}

\noindent where $q$ is a linear query that may be randomized and $f$ is a function from an entry to an element in an Abelian group, e.g. a scalar or a fixed array.  $E[q(\mathbf{x})]$ is linear to the number of occurrences of an entry in a database.

When $f(x) = x$, the query is the sum of entries. When $f(x) = \text{IF}(cond(x), 1, 0)$, the query is a conditional counting. When $f(x) = x^2$, the query is the sum of second moments.

$E[q(\mathbf{x})]$ changes when $\mathbf{x}$ changes with time. The change of $E[q(\mathbf{x})]$ only depends on the change of $\mathbf{x}$ during a time period:

\begin{equation}
    \Delta E[q(\mathbf{x})] =  E[\Delta q(\mathbf{x})] = \sum_{x \in \mathbf{x}} \Delta  f(x).
\end{equation}

\noindent Also, the change in expectation is additive:

\begin{equation}
    E[q(\mathbf{x})] |_{t_1}^{t_3} = \sum_{x \in \mathbf{x}} f(x) |_{t_1}^{t_2} + \sum_{x \in \mathbf{x}} f(x) |_{t_2}^{t_3},
    \label{eq:linear_query_additive}
\end{equation}

\noindent where $x|_{a}^b$ represents the change in $x$ from $a$ to $b$. $\Delta q(\mathbf{x})$ is referred to as linear-query change. Given Eq. \eqref{eq:linear_query_additive}, the continual release of $\Delta q(\mathbf{x})$ could continually estimate the continual release of $q(\mathbf{x})$ over time.

% Therefore, linear-query changes against a dynamic database can be continually released and each change only depends on the change of the database from the last release. The results of the continual release can be continually summed to derive a continual release of the linear query against the database. 

\section{Changelog Representation of Dynamic Databases}

This section discusses how to represent a dynamic database as a changelog, and how a query is run against a changelog.

\subsection{Changelog Definition}
\label{sec:changelog_definition}

A dynamic database can be represented by a changelog, which contains all the changes of a database. The entries of a changelog are immutable once they are committed to the changelog. 

The entries of a changelog are called mutations, each of which records the change of an entry at a timestamp. The mutation includes the insertion and deletion of the entry, and the modification of its value. The mutation of an entry $x$ at time $t$ is denoted as $m_{x, t}$. Usually, a mutation exists only if the entry has some change. However, for convenience, we may use $m_{x, t} = \text{null}$ to denote no mutation of the entry $x$ at time $t$. The subscript of $m_{x, t}$ may be dropped (e.g. $m$) if there is no ambiguity. A sorted sequence of $m$ is denoted as $\mathbf{m}$, where mutations are first sorted by $t$ and then by $x$.

Define the initial state of a database $\mathbf{x}$ to be $\mathbf{x}_{t_0}$, where $t_0$ is the creation time of the database. The state of $\mathbf{x}$ at any time $t$ can be represented by $\mathbf{x}_{t_0} + \mathbf{m}_{\mathbf{x}, (t_0, t]}$, where "+" represents applying a sequence of mutations toward a database, and $\mathbf{m}_{\mathbf{x}, (t_0, t]}$ is the sorted sequence of $\{ m_{x, t'} | x \in \mathbf{x}, t_0 < t' \le t, m_{x, t'} \ne \text{null}  \}$.

The creation of a database is equivalent to simultaneously inserting the initial entries of the database at $t_0$, and $\mathbf{x}_{t_0}$ can be represented by a sequence of mutations $\mathbf{m}_{\mathbf{x}, t_0} = \{ m_{x, t_0} | x \in \mathbf{x}_{t_0}, m_{x, t_0} = \text{insert } x \}$. Thus, an arbitrary state of $\mathbf{x}_t $ can be represented by a sequence of mutations $\mathbf{m}_{\mathbf{x}, t} = \mathbf{m}_{\mathbf{x}, t_0} + \mathbf{m}_{\mathbf{x}, (t_0, t]}$. $\mathbf{m_x}$ denotes all the mutations during the lifetime of a database $\mathbf{x}$, which is also called the changelog of the database. Similarly, $\mathbf{m}_x$ denotes all the mutations of an entry $x$.

\subsection{Mutation}

There is no requirement for the format of a mutation, as long as "null" represents no mutation. In practice, $m$ can store a new value or the change of a value. For example, if an entry is updated from 50 to 100, the "new value" format will be $m = 100$ and the "value change" format will be $m = 50 \rightarrow 100$. Any format is allowed as long as the query on mutations can satisfy DP requirements. However, the "value change" format is preferred and will be used in this article as it provides all the information of a change.

\subsection{Query on Mutations}

A mutation query $q: \mathbb{M}^\mathbb{N} \times \mathbb{R}$ takes a sequence of mutations as input and outputs a randomized value in range $\mathbb{R}$. In this article, $q$ is further broken down into a mechanism $M$ and a mutation filter $d$, such that $q(\cdot) = M(d(\cdot))$. The mechanism $M$ is a more generic query mapping from a sequence of mutations to a randomized output. The filter $d$ filters the input mutations based on a predicate. The predicate is predefined before being against the mutations. Thus, $d$ represents a subset of the mutation universe $\mathbb{M}$. $m \in d$ denotes that a mutation $m$ is accepted by the predicate of a filter $d$. A type of filter frequently used in this article is a time-range filter, which is represented by a time range, e.g. $(a, b]$, and accepts the mutations occurring in the given time range.

\section{Global Differential Privacy}

This section studies how continual releases against a dynamic database could achieve global differential privacy.

\subsection{Adjacent Changelogs}

Suppose two dynamic databases, $\mathbf{x}$ and $\mathbf{x}'$, differ by one entry $x$, and assume $\mathbf{x}$ + $x$ = $\mathbf{x}'$. Their changelogs have the following relationship:

\begin{equation}
    \mathbf{m_x} + \mathbf{m}_x = \mathbf{m_{x'}}, 
\end{equation}

\noindent where $\mathbf{m}_x$ denotes all the mutations of the entry $x$, and "+" denotes inserting mutations and maintaining the sorted order of the changelogs. Any pair of changelogs satisfying the above relationship is called adjacent.

\subsection{Differential Privacy}

Given a continual release comprising a stream of queries $\mathbf{q} = (q_i)_{i \in [n]^{+}}$, where $n \in \mathbb{N}^{+}$. $q_i(\cdot) = M_i(d_i(\cdot))$ and $(\epsilon_i, \delta_i)$-DP. $\mathbf{q}$ randomly outputs a result $\mathbf{r} = (r_i)_{i \in [n]^{+}}$. Assume the result of a query does not depend on the results of its previous queries, the probability of outputting $\mathbf{r}$ is

\begin{equation}
   Pr(\mathbf{q(\mathbf{m})} = \mathbf{r}) = \prod_{i \in [n]^{+}} Pr[M_i(d_i(\mathbf{m})) = r_i], 
\end{equation}

\noindent where $\mathbf{m}$ is the changelog of a database.

Suppose $\mathbf{m}$ and $\mathbf{m}'$ are two adjacent changelogs. If a continual release $\mathbf{q}$ is $(\epsilon, \delta)$-DP, for any subset $S \subseteq \mathbb{R}^n$, it must satisfy:

\begin{multline}
\sum_{\mathbf{r} \in S} \prod_{i \in [n]^{+}} Pr[M_i(d_i(\mathbf{m})) = r_i] \le \\
e^{\epsilon} \left( \sum_{\mathbf{r} \in S} \prod_{i \in [n]^{+}} Pr[M_i(d_i(\mathbf{m}')) = r_i] \right) + \delta .
\label{eq:continual_release_dp}
\end{multline}

Now we will derive $(\epsilon, \delta)$ from $(\epsilon_i, \delta_i)$. Define $\mathbf{q} | m = \{q_i | q_i \in \mathbf{q}, m \in d_i \}$ to be a group of queries in $\mathbf{q}$ whose filters accept a mutation $m$. Furthermore, we can define 

\begin{equation}
    \mathbf{q} | x = \underset{m \in \mathbf{m}_x}{\cup} \mathbf{q} | m,
    \label{eq:q_affected_by_x}
\end{equation}

\noindent where $x$ is an entry. If $q \in \mathbf{q} | x$, we call $q$ is affected by $x$. We can see $q_i \in \mathbf{q} | x$ is equivalent to $x \in d_i$ in Eq. \eqref{eq:L_q_x}, so we have

\begin{equation}
  L(\mathbf{q}, x) = \underset{q_i \in \mathbf{q} | x}{SC} (\epsilon_i, \delta_i),   
  \label{eq:parallel_composition_of_mutations}
\end{equation}

\noindent and Theorem \ref{thm:parallel_composition} proves that

\begin{equation}
    (\epsilon, \delta) =  \sup_{x \in \mathcal{X}} \: \underset{q_i \in \mathbf{q} | x}{SC} (\epsilon_i, \delta_i). 
\end{equation}

Suppose an entry $x$ mutates at every timestamp. It will affect every query of a continual release, so privacy loss of the continual release will be $SC_{q_i \in \mathbf{q}} (\epsilon_i, \delta_i)$, which is not unbounded as it increases with the number of queries. The following sections will introduce two constraints that can provide the continual release of a practical privacy-loss bound.

\subsection{Constraints on Mutations}
% In order to bound the differential privacy loss of a dynamic database, its entries should obey some constraints. This article will introduce two kinds of constraints in the coming sections.

\subsubsection{At-most-k Mutations}

If the states of an entry can be represented by a finite-state machine with a directed acyclic graph (DAG) as its transition diagram, then there will be a longest path within the diagram, which limits the maximum number of mutations that can occur for that entry. In cases where the state-transition diagram of an entry is not a finite-state DAG, we can still group the states into composite states that form a finite-state DAG. Firstly, if an entry has a cyclic transition diagram, the strongly connected components (SCCs) of the diagram still form a DAG. Thus, a SCC can be considered as a composite state. Secondly, if an entry has infinite states, we can group them into a finite number of composite states. For instance, a counter that can mutate from zero to infinity can be grouped into zero, one, two, and beyond.

If every entry in a database has at-most $k$ mutations, and each mutation can only impact a finite number of queries, then the entry can only affect a finite number of queries. Consequently, regardless of the number of queries performed by a continual release against the database, the privacy loss is bounded.

\subsubsection{Time-bounded Mutations}

Some databases allow their entries to be modified for a period of time after they have been inserted. Here are two common scenarios: (1) any objects with a lifetime, e.g. a monthly pass, the admission status of a student; (2) a database has multiple data sources, where one source may overwrite the data from another source. For instance, in the Lambda Architecture for Big Data, a database is first written with some real-time but inaccurate data, and then the source-of-the-truth batch data is written later to overwrite the real-time data. The batch data is written every day, so an entry of the database will not be modified after one day.

\begin{definition}
    If a database is guaranteed to have its entries modified within a period of time with length of $B$ after they have been inserted, then the database is defined to be $B$-time-bounded in mutations.
\end{definition}

\subsection{Disjoint Continual Release}

Let's first define "disjoint". A query filter accepts a subset of the mutation universe $\mathbb{M}$. Given two query filters $d_i$ and $d_j$, define $d_i \cap d_j = \{ m \in \mathbb{M} | m \in d_i, m \in d_j \}$.

\begin{definition}
    (Disjoint) $d_i$ and $d_j$ are disjoint iff $d_i \cap d_j = \emptyset$.
\end{definition}

\noindent Now we can define a disjoint continual release. 

\begin{figure}[htp!]
\centerline{\includegraphics[width=1.0\columnwidth]{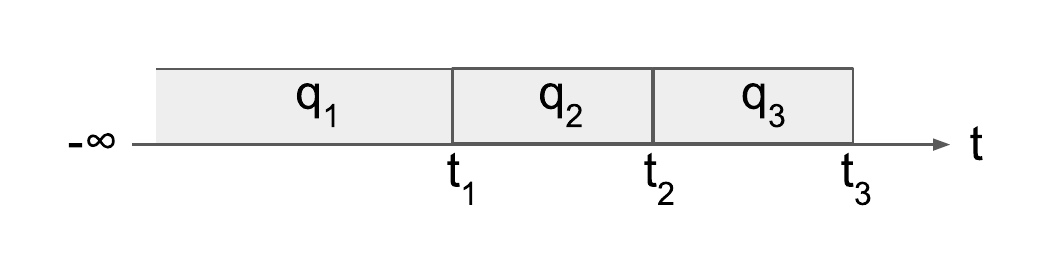}}
\caption{Disjoint Continual Release}
\label{fig: DCR}
\end{figure}

\begin{definition}
\label{def:DCR}
    If a continual release $\mathbf{q}$ satisfies the following conditions:

1. The first query $q_1$ has a time-range filter $(-\infty, t_1]$;

2. An arbitrary query $q_i$ has a time-range filter $(t_{i-1}, t_i]$;

3. all queries are $(\epsilon, \delta)$-DP,

\noindent then $\mathbf{q}$ is a disjoint continual release (Fig. \ref{fig: DCR}). 
\end{definition}

$\text{DCR}(\epsilon, \delta, \mathbf{t})$ denotes the collection of the disjoint continual releases satisfying the above requirements, where $\mathbf{t} =(t_1, t_2, ...)$. $\text{DCR}(\epsilon, \delta, \mathbf{t}, M)$ denotes a continual release whose queries have the same mechanism $M$. It is trivial to realize that a DCR is disjoint because any pairs of its queries have disjoint filters. 

\subsubsection{Linear-query Change Continual Release}

As discussed in Section \ref{sec:linear_query}, a linear-query change ($\Delta LQ$) continual release exports the changes of a linear query, and each query only reads the mutations since its last query. Thus, the linear-query change continual release belongs to DCRs. It can be represented by $\text{DCR}(\epsilon, \delta, \mathbf{t}, \Delta LQ)$, where $t_i \in \mathbf{t}$ is the execution time of the corresponding query $q_i$, and the mechanism $\Delta LQ$ is a function like:

{\tt \small
\begin{verbatim}
def linear_query_change(mutations):
    change = 0
    for mutation in mutations:
        change -= f(mutation.prev_value)
        change += f(mutation.current_value)
    return change
\end{verbatim}
}

\noindent where $f$ represents the $f(x)$ in the Eq. \eqref{eq:linear_query}. Suppose $f(x)$ has a bounded range, i.e. $f(x) \in [a, b]$. Then $\Delta LQ$ has a range of $[a - b, b - a]$. If Laplace or Gaussian Mechanism is used to ensure differential privacy, the $L_2$ sensitivity of $\Delta LQ$ used by the mechanisms will be $|b - a|$. If the range of $f(x)$ is unbounded, it might be truncated to a bounded range without losing too much accuracy \cite{perrier2018PAK, wang2021continuous}.

\subsubsection{At-most-k Mutations}

\begin{theorem}
    If $\mathbf{q} \in \text{DCR}(\epsilon, \delta, \mathbf{t})$ against a dynamic database whose entries are mutated at most $k$ times, $\mathbf{q}$ is $k \text{-} (\epsilon, \delta)$-DP.
    \label{thm:DCR_DP_at_most_k}
\end{theorem}

\noindent Proof: the query filters of $\mathbf{q}$ are disjoint, so a mutation can only affect at most one query of the continual release. An entry $x$ has at most $k$ mutations, so it can at most affect $k$ queries, i.e. $|\mathbf{q} | x| = k$. Since all the queries have the same privacy loss, Eq. \eqref{eq:parallel_composition_of_mutations} becomes

\begin{equation}
     L(\mathbf{q}, x) = \underset{[k]^{+}}{SC} \; (\epsilon, \delta) = k \text{-} (\epsilon, \delta).
\end{equation}

\noindent $L(\mathbf{q}, x)$ is independent of $x$. Given Theorem \ref{thm:parallel_composition}, $\mathbf{q}$ is $k\textbf{-}(\epsilon, \delta)$-DP. 

\subsubsection{Time-bounded Mutations}

\begin{figure}[htp!]
\centerline{\includegraphics[width=1.0\columnwidth]{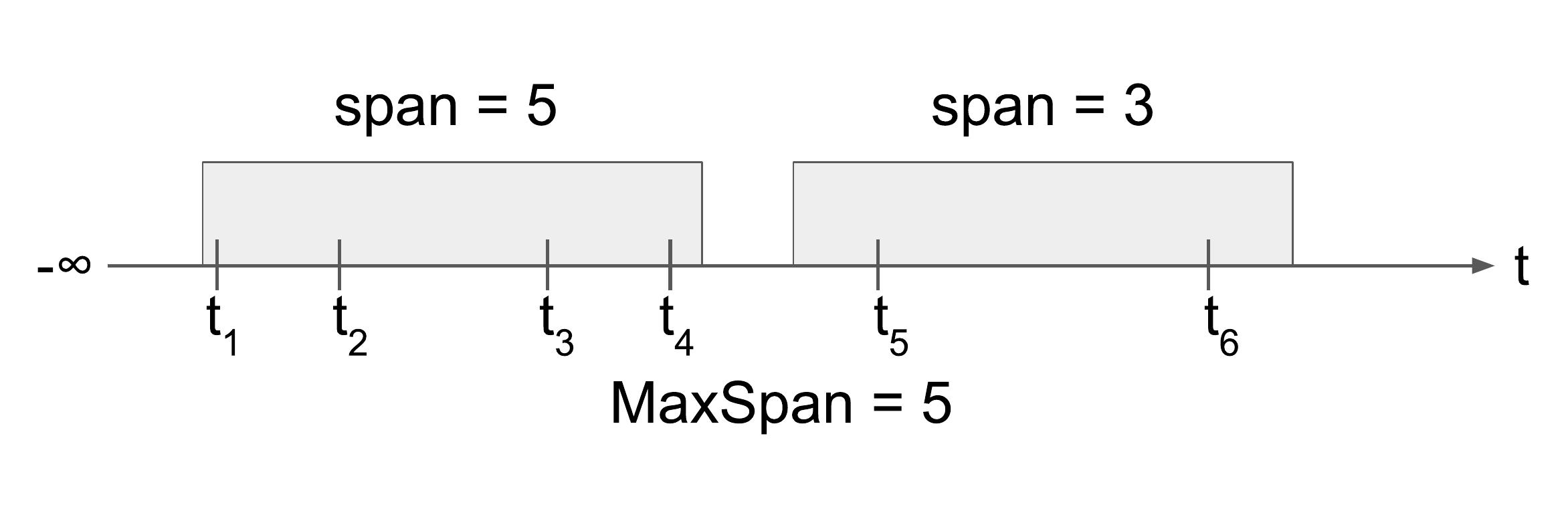}}
\caption{MaxSpan illustration}
\label{fig: max_span}
\end{figure}

\begin{definition}
    $ \text{MostSpan}(\mathbf{t}, T)$ is a function that returns maximum numbers of consecutive time ranges $(t_{i-1}, t_i]$ that are overlapped with a sliding window of length $T$ (Fig. \ref{fig: max_span}).
\end{definition}

\noindent One implementation is that
{\tt \small
\begin{verbatim}
def MostSpan(t_array, T):
    res = 0
    for i in range(0, len(t_array) - 1):
        for j in range(i + 1, len(t_array)):
            if t_array[j] - t_array[i] >= T:
                res = max(res, j - i + 1)
                break
    return res
\end{verbatim}
}

\begin{theorem}
     If $\mathbf{q} \in \text{DCR}(\epsilon, \delta, \mathbf{t})$ queries against a dynamic database being $B$-time-bounded in mutations, $\mathbf{q}$ is $\text{MostSpan}( \mathbf{t}, B) \text{-} (\epsilon, \delta)$-DP.
     \label{thm:DCR_DP_time_bounded}
\end{theorem}

\noindent Proof: the mutations of an entry can only occur within a period of length $B$, and affect at most $\text{MostSpan}(\mathbf{t}, B)$ queries of $\mathbf{q}$. That is, $|\mathbf{q} | x| \le \text{MostSpan}( \mathbf{t}, B)$ for any entries. Similar to above, the privacy loss of $\mathbf{q}$ is thus $\text{MostSpan}(\mathbf{t}, B) \text{-} (\epsilon, \delta)$-DP.

In practice, a DCR usually has a constant interval between two queries, i.e. $t_i - t_{i-1} = W$, where $W$ is a constant. The endpoints of its filters $\mathbf{t} = t_1 + W[n]$, where $n$ is the length of the DCR.

\begin{corollary}
    $\text{DCR}(\epsilon, \delta, t_1 + W[n])$ is $(\lceil \frac{B}{W} \rceil + 1) \text{-} (\epsilon, \delta)$-DP if it queries against a database being $B$-time-bounded in mutations.
\end{corollary}

\noindent Proof: all the query filters have the same interval $W$. At most $\lceil \frac{B}{W} \rceil + 1$ of them can overlap with a sliding window of size $B$, so $\text{MostSpan}( \mathbf{t}, B) = \lceil \frac{B}{W} \rceil + 1$.

\subsubsection{Hierarchical Disjoint Continual Release} 

Previous literature \cite{dwork2010continual, chan2011private, qardaji2013} proposes an efficient algorithm to derive a continual release of aggregatable queries from a hierarchy of disjoint continual releases. A query $q$ comprises a mechanism $M$ and a filter $d$.

\begin{definition}
    (Aggregatable Query) $q$ is aggregatable if, for any set of disjoint query filters $\mathbf{d}' = (d_i')_{i \in [n]^{+}}$ where $ \cup_{i \in [n]^{+}} d_i' = d$, there exists an aggregation operator $A: \mathbb{R}^n \rightarrow \mathbb{R}$ that satisfies

\begin{equation}
    E[q(\cdot)] = E[A \left( M(d'_1(\cdot)), M(d'_2(\cdot)), ..., M(d'_i(\cdot)), ... \right) ].
    \label{eq:aggregatable}
\end{equation}

\end{definition}

Now we can introduce the hierarchical disjoint continual release (HDCR). A HDCR comprises layers of disjoint continual releases. The interval between queries increases exponentially from the bottom layer to the top one. Let the length of the HDCR be $T$ (a duration). The HDCR starts at $t_s$. 

\begin{definition}
\label{def:HDCR}
    $\text{HDCR}(\epsilon, \delta, h, c, t_s, T, W, M)$ denotes the following hierarchy of disjoint continual releases:
\begin{multline}
\text{HDCR}(\epsilon, \delta, h, c, t_s, T, W, M) = \\
     \begin{cases}
      \text{layer } h-1 \text{: DCR}(\epsilon, \delta, t_s + c^{h-1} W[\lceil \frac{T}{c^{h-1} W} \rceil], M)\\
      ...\\
      \text{layer } i \text{: DCR}(\epsilon, \delta, t_s + c^i W[\lceil \frac{T}{c^i W} \rceil], M) \\
      ...\\
      \text{layer 0: DCR}(\epsilon, \delta, t_s + W[\lceil \frac{T}{W} \rceil], M)
    \end{cases},
\end{multline}

\noindent where $h$ is the height of the HDCR; $c$ is a constant branching factor; $W$ is the interval size of the bottom continual release; and $M$ is a query mechanism. 
\end{definition}

\noindent The queries of each layer of the HDCR are called the nodes of the HDCR.

\begin{theorem}
    $\text{HDCR}(\epsilon, \delta, h, c, t_s, T, W, M)$ is $hk \text{-} (\epsilon, \delta)$-DP for a database with at-most-k mutations or $\left[ \sum_{i \in [h-1]}(\frac{1}{c^i} \lceil \frac{B}{W} \rceil + 1) \right] \text{-} (\epsilon, \delta)$-DP for a database with $B$-time-bounded mutations.
\end{theorem}

\noindent Proof: a HDCR composes $h$ layers of $(\epsilon, \delta)$-DP DCR against the same changelog of a database, so its privacy loss follows the sequential composition (Corollary \ref{cor:nested_k_fold}).

\cite{dwork2010continual, chan2011private, qardaji2013} demonstrated any linear queries at $t \le T$ can be computed from $O(\log T)$ nodes of a hierarchical algorithm. This paper extends this result to the aggregatable queries derived from a HDCR. Suppose that a query $q$ has a time-range filter $(l W, r W]$, where $l, r \in \mathbb{N}$, and is aggregatable from the queries with a mechanism $M$. Without losing generality, set the start time $t_s$ of a HDCR to zero and the length $T$ to be larger than $r W$. 

\begin{theorem}
    $q$ can be aggregated from at most $2 (c - 1) h_{min}$ nodes of $\text{HDCR}(\epsilon, \delta, h, c, 0, T, W, M)$ if $h \ge h_{min}$, where $h_{min} = \lceil \log_c(r - l) \rceil$ .
    \label{thm:aggregatable_from_HDCR}
\end{theorem}

\noindent The proof is at Appendix \ref{sec:aggregatable_from_HDCR}. \cite{qardaji2013} implies the average number of nodes to aggregate is $O((c-1)h_{min})$. \cite{chan2011private} finds that at most $O(h_{min})$ nodes are needed to aggregate when $c=2$. Both are consistent with Theorem \ref{thm:aggregatable_from_HDCR}.

Therefore, the result of a linear query change is equivalent to the sum of the results from at most $2 (c - 1) h_{min}$ nodes of the HDCR. Define $\sigma^2$ to be the variance of the nodes, and we have

\begin{corollary}
\label{cor:HDCR_var}
    $q$ has a variance of $2 (c - 1) h_{min} \sigma^2$, if $q$ has a mechanism of $\Delta LQ$.
\end{corollary}

\subsection{Sliding Window Continual Release} 

Data analysts often continually query the change of a database in a certain period. For example, researchers may be interested in a continual release of how the positive cases of a disease change in the last 14 days. 

\begin{figure}[htp!]
\centerline{\includegraphics[width=1.0\columnwidth]{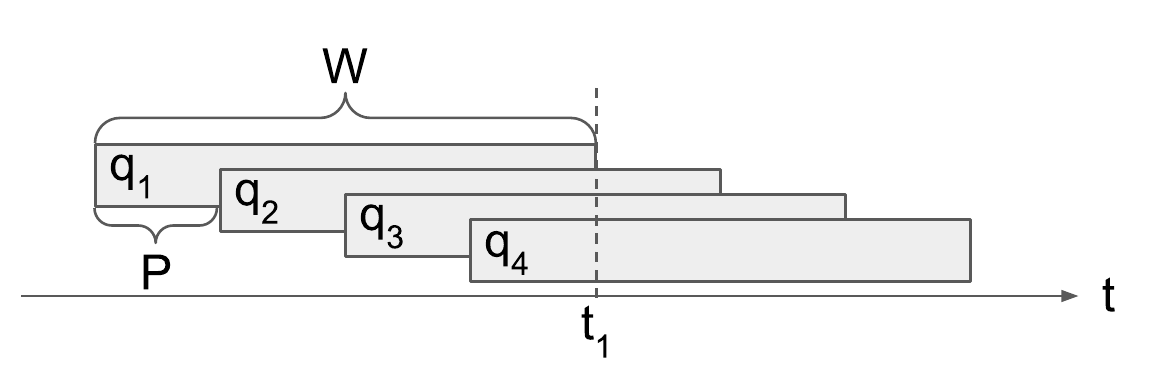}}
\caption{Sliding-window Continual Release}
\label{fig: SWCR}
\end{figure}

\begin{definition}
\label{def:SWCR}
    A sliding-window continual release (Fig. \ref{fig: SWCR}) is a stream of queries with the following properties:

1. for any query $q_i$, it has a time-range filter $(t_i - W, t_i]$, where the window size $W$ is constant;

2. any pair of consecutive queries $q_i$ and $q_{i+1}$ have $t_{i+1} - t_i = P$, where the period $P$ is a constant;

3. all queries are $(\epsilon, \delta)$-DP.

\end{definition}

$\text{SWCR}( \epsilon, \delta, P, W, t_1)$ denotes the collection of continual release with the above requirements. If all the queries in a SWCR have the same mechanism $M$, $\text{SWCR}( \epsilon, \delta, P, W, t_1, M)$ denotes such a continual release.

\subsubsection{At-most-k Mutations}

\begin{theorem}
    If $\mathbf{q} \in \text{SWCR}( \epsilon, \delta, P, W, t_1)$ is queried against a dynamic database whose entries are mutated at most $k$ times, $\mathbf{q}$ is $k \text{-} \lceil \frac{W}{P} \rceil \text{-} (\epsilon, \delta)$-DP.
    \label{thm:SWCR_DP_at_most_k}
\end{theorem}

\noindent Proof: one mutation occurring at $t$ can overlap with the queries whose filters have right ends in $[t, t + W)$. At most $\lceil \frac{W}{P} \rceil$ right ends can fit into this range because any pair of consecutive queries $q_i$ and $q_{i+1}$ should have $t_{i+1} - t_i = P$. Since one entry has at most $k$ mutations, the affected queries are at most $k \lceil \frac{W}{P} \rceil$, i.e. $|\mathbf{q} | x| \le k \lceil \frac{W}{P} \rceil$. Thus, the privacy loss of $\mathbf{q}$ has a upper bound of $k \text{-} \lceil \frac{W}{P} \rceil \text{-} (\epsilon, \delta)$.

\subsubsection{Time-bounded Mutations}

\begin{theorem}
    If $\mathbf{q} \in \text{SWCR}( \epsilon, \delta, P, W, t_1)$ is queried against a dynamic database being $B$-time-bounded in mutations, $\mathbf{q}$ is $\lceil \frac{B + W}{P} \rceil \text{-} (\epsilon, \delta)$-DP.
    \label{thm:SWCR_DP_time_bounded}
\end{theorem}

\noindent Proof: an arbitrary entry mutates at $[t, t + B]$, which overlaps with the queries whose filters have right ends in $[t, t + B + W)$. Similar to the above section, at most $\lceil \frac{B + W}{P} \rceil$ right ends can fit into this range, and one entry can affect at most $\lceil \frac{B + W}{P} \rceil$ queries. Therefore, the $\mathbf{q}$ is $\lceil \frac{B + W}{P} \rceil \text{-} (\epsilon, \delta)$-DP.

\subsubsection{Convert to the Hierarchical Disjoint Continual Release}

\begin{figure}[htp!]
\centerline{\includegraphics[width=1.0\columnwidth]{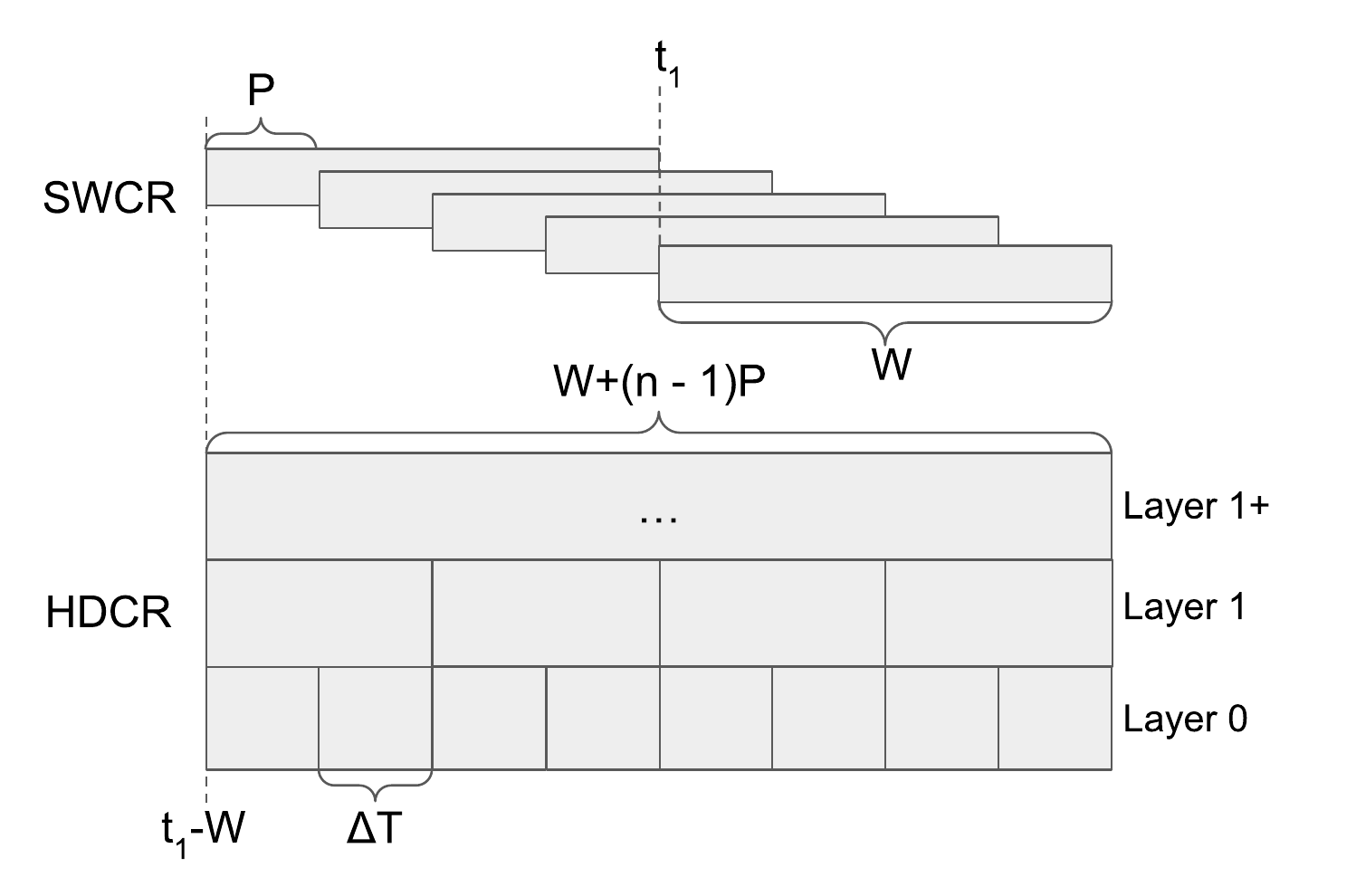}}
\caption{Derive SWCR from HDCR}
\label{fig: HDCR_to_SWCR}
\end{figure}

If the queries of a SWCR are aggregatable, the SWCR can be obtained from the aggregations from a HDCR whose bottom layer has an interval size equivalent to the greatest common divisor of $W$ and $P$ of the SWCR (Fig. \ref{fig: HDCR_to_SWCR}). Define $\Delta T = GCD(W,P)$.

Formally, suppose a sliding-window continual release $\mathbf{q} \in \text{SWCR}( \epsilon, \delta, P, W, t_1)$ has a size of $n$ and is aggregatable from mechanism $M$. The start time of the HDCR is $t_1 - W$, and the timespan of the HDCR equals the timespan of the SWCR plus $W$, i.e. $W + (n - 1) P$. Given Theorem \ref{thm:aggregatable_from_HDCR}, the $h$ of the HDCR should be at least $\lceil \log_c \frac{W}{\Delta T} \rceil$ to constrain the variance to a logarithm complexity. Define each node of the HDCR to be $(\epsilon', \delta')$-DP. Now, we can claim 

\begin{theorem}
     $\text{SWCR}( \epsilon, \delta, P, W, t_1)$ can be aggregated from the nodes of $\text{HDCR}(\epsilon', \delta', \lceil \log_c \frac{W}{\Delta T} \rceil, c, t_1 - W, W + (n - 1) P, \Delta T, M)$. 
\end{theorem}

Suppose the continual release above is a linear query change whose mechanism $\Delta LQ$ has a variance of $O(\frac{1}{\epsilon^2})$, e.g. Laplace Mechanism. Also only consider $\epsilon$-DP. Now, we like to compare the variance of the SWCR and the corresponding HDCR with the same privacy loss. When querying against a database whose entries mutate at most $k$ times, the privacy loss of the HDCR and SWCR are $k \lceil \log_c \frac{W}{\Delta T} \rceil \epsilon'$ and $k \lceil \frac{W}{P} \rceil \epsilon$, respectively. With the same privacy loss, we have $\epsilon' = \lceil \frac{W}{P} \rceil \frac{\epsilon}{\lceil \log_c \frac{W}{\Delta T} \rceil}$. 

\begin{theorem} \label{thm:HDCR_vs_SWCR_at_most_k}
$\text{HDCR}(\lceil \frac{W}{P} \rceil \frac{\epsilon}{\lceil \log_c \frac{W}{\Delta T} \rceil}, 0, \lceil \log_c \frac{W}{\Delta T} \rceil, c, t_1 - W, W + (n - 1) P, \Delta T, \Delta LQ)$ derives the result of $\text{SWCR}( \epsilon, 0, P, W, t_1, \Delta LQ)$ with a lower variance, if they query against a database whose entries mutate at most $k$ times and 

\begin{equation}
    2(c-1) (\lceil \log_c \frac{W}{\Delta T} \rceil)^3 < \lceil \frac{W}{P} \rceil^2.
    \label{eq:HDCR_vs_SWCR_at_most_k}
\end{equation}
\end{theorem}

\noindent The proof is at Appendix \ref{sec:proof_of_HDCD_vs_SWCR}. 

When querying against a database being $B$-time-bounded in mutations, the privacy loss of the HDCR and SWCR are $[\frac{c^h - 1}{c^h - c^{h-1}} \lceil \frac{B}{\Delta T} \rceil + h] \epsilon'$ and $\lceil \frac{W + B}{P} \rceil \epsilon$, respectively, where $h = \lceil \log_c \frac{W}{\Delta T} \rceil$. With the same privacy loss, we have $\epsilon' = \frac{\lceil \frac{W + B}{P} \rceil}{\frac{c^h - 1}{c^h - c^{h-1}} \lceil \frac{B}{\Delta T} \rceil + h} \epsilon$.

\begin{theorem}
$\text{HDCR}(\frac{\lceil \frac{W + B}{P} \rceil}{\frac{c^h - 1}{c^h - c^{h-1}} \lceil \frac{B}{\Delta T} \rceil + h} \epsilon, 0, h, c, t_1 - W, W + (n - 1) P, \Delta T, \Delta LQ)$ derives the result of $SWCR( \epsilon, 0, P, W, \\ t_1, \Delta LQ)$ with a lower variance, if they query against a database  being $B$-time-bounded in mutations and the following inequality holds:
\begin{equation}
2(c - 1) h (\frac{c^h - 1}{c^h - c^{h-1}} \lceil \frac{B}{\Delta T} \rceil + h)^2 < \lceil \frac{W + B}{P} \rceil^2,
\label{eq:HDCR_vs_SWCR_time_bounded}
\end{equation}

\noindent where $h = \lceil \log_c \frac{W}{\Delta T} \rceil$.

\label{thm:HDCR_vs_SWCR_time_bounded}
\end{theorem}

\noindent The proof is at Appendix \ref{sec:proof_of_HDCD_vs_SWCR}.

\subsection{Database with Hybrid Constraints on Mutations}

Think about a database recording the tickets of public transportation in a city. There are two types of fares: multiple-ride ticket and monthly pass. A multiple-ride ticket can only be used a limited number of times, which is equivalent to at-most $k$ mutations, while a monthly pass can be used unlimited times in a certain period, i.e. time-bounded mutations. Therefore, it is necessary to study the differential privacy of a database whose entries have various constraints on mutations.

Let $\mathbf{G}$ be a list of constraints on mutations. $\mathbf{G} = (G_1, G_2, ..., G_g)$, where $G_i$ is the $i$-th constraint. Suppose there exists a continual release $\mathbf{q}$ such that, for an arbitrary constraints $G_i \in \mathbf{G}$, it is $(\epsilon_i, \delta_i)$-DP against a database with the constraint $G_i$.

\begin{theorem}
    $\mathbf{q}$ is $[\sup_{i \in [g]^{+}} (\epsilon_i, \delta_i)]$-DP against a database whose entries are constrained by any $G \in \mathbf{G}$.
\end{theorem}

\noindent Proof: Define $\mathbf{x}$ to be such a database. If an entry $x \in \mathbf{x}$ is constrained by $G_i$, Eq. \eqref{eq:parallel_composition_of_mutations} becomes

\begin{equation}
    L(\mathbf{q}, x) = (\epsilon_i, \delta_i).
\end{equation}

\noindent Since an entry $x$ can be constrained by any one of $\mathbf{G}$, given Theorem \ref{thm:parallel_composition}, $\mathbf{q}$ is $[\sup_{i \in [g]^{+}} (\epsilon_i, \delta_i)]$-DP.

\section{Local Differential Privacy}

\subsection{Changelog Representation of Individual Entry}

Similar to Section \ref{sec:changelog_definition}, The evolution of entry $x$ can also be represented a changelog $\mathbf{m}_x$, which is defined to be $\{m_{x, t} | t \in \mathbb{T}, m_{x, t} \ne \text{null} \}$. Similar to the query against a database, a query against an entry also has a filter and mechanism. If no mutations of an entry satisfy the filter, the query will receive an empty set of mutations as input. Similarly, if the time-range filter of a query is outside the lifetime of an entry, it will receive an empty set as well.

\subsection{Differential Privacy}
\label{sec:local_dp}

For any two entries $x, x' \in \mathcal{X}$, their changelogs are $\mathbf{m}$ and $\mathbf{m}'$, respectively. A query $q$ is local $(\epsilon, \delta)$-DP iff for any $S \subseteq \mathbb{R}$

\begin{equation}
Pr(q(\mathbf{m}) \in S) \le e^{\epsilon} Pr(q(\mathbf{m}') \in S) + \delta. 
\end{equation}

\subsection{Continual Release}

A continual release against an entry is the same as that against a database, except that the input is the changelog of an entry. Define $\mathbf{q} = (q_i)_{i \in [n]^{+}} $. A query $q_i \in \mathbf{q}$ has a filter $d_i$, and is $(\epsilon_i, \delta_i)$-DP. Define $\mathbf{q} | (\mathbf{m} \oplus \mathbf{m}') = \{q_i \in \mathbf{q} \; | \; d_i(\mathbf{m}) \ne d_i(\mathbf{m}')  \}$, and 

\begin{equation}
    L(\mathbf{q}, \mathbf{m}, \mathbf{m}') = (\epsilon_{\mathbf{m}, \mathbf{m}'}, \delta_{\mathbf{m}, \mathbf{m}'}) = \underset{q_i \in \mathbf{q} | (\mathbf{m} \oplus \mathbf{m}')}{SC} (\epsilon_i, \delta_i).
\end{equation}

\noindent If $d_i(\mathbf{m}) = d_i(\mathbf{m}')$, then $q_i$ will receive identical inputs and generate the same output distribution. Similar to Eq. \eqref{eq:parallel_composition_step_1} - \eqref{eq:parallel_composition_step_3}, we have

\begin{equation}
     \prod_{i} Pr[q_i(\mathbf{m})] \le e^{\epsilon_{\mathbf{m}, \mathbf{m}'}} \prod_{i} Pr[q_i(\mathbf{m}')] + \delta_{\mathbf{m}, \mathbf{m}'}.
\end{equation}

\noindent Similar to Theorem \ref{thm:parallel_composition}, $\mathbf{q}$ is $[\sup_{\mathbf{m}, \mathbf{m}'} L(\mathbf{q}, \mathbf{m}, \mathbf{m}')]$-DP. 

\begin{theorem}
    $\mathbf{q}$ is also $[2 \text{-} \sup_{\mathbf{m}} L(\mathbf{q}, \mathbf{m})]$-DP.
    \label{thm:local_dp_comp}
\end{theorem}

\noindent Proof: Define $\mathbf{q} | \mathbf{m} = \{ q_i \in \mathbf{q} | d_i(\mathbf{m}) \ne \text{null} \}$, $\mathbf{q} | (\mathbf{m} \cup \mathbf{m}')  = \{ q_i \in \mathbf{q} | d_i(\mathbf{m}) \ne \text{null} \text{ or } d_i(\mathbf{m}') \ne \text{null} \}$, and $\mathbf{q} | \mathbf{m} + \mathbf{q} | \mathbf{m}'$ to be joining two sets without deduplication. By definition, we have 

\begin{equation}
    \mathbf{q} | (\mathbf{m} \oplus \mathbf{m}') \subseteq \mathbf{q} | (\mathbf{m} \cup \mathbf{m}') \subseteq \mathbf{q} | \mathbf{m} + \mathbf{q} | \mathbf{m}' .
\end{equation}

\noindent Also, we define

\begin{equation}
    L(\mathbf{q}, \mathbf{m}) = \underset{q_i \in \mathbf{q} | \mathbf{m}}{SC} (\epsilon_i, \delta_i).
\end{equation}

\noindent Since the composition of differential privacy is monotonically increasing  \cite{dwork2014privacy}, i.e. $(\epsilon, \delta) \preceq SC[(\epsilon, \delta), (\epsilon', \delta')]$, we have

\begin{equation}
    L(\mathbf{q}, \mathbf{m} , \mathbf{m}') \preceq SC[  L(\mathbf{q}, \mathbf{m}), L(\mathbf{q}, \mathbf{m}')] .
\end{equation}

\noindent We can extend this to any pair of changelogs., 
\begin{multline}
        \sup_{\mathbf{m}, \mathbf{m}'} L(\mathbf{q}, \mathbf{m}, \mathbf{m}') \preceq  SC[\sup_{\mathbf{m}} L(\mathbf{q}, \mathbf{m}), \sup_{\mathbf{m}'}L(\mathbf{q}, \mathbf{m}')] \\ = 2 \text{-} \sup_{\mathbf{m}} L(\mathbf{q}, \mathbf{m}) . \qed
\end{multline}

Notice the definition of $L(\mathbf{q}, \mathbf{m})$ in local DP is identical to $L(\mathbf{q},  x)$ in global DP. With Theorem \ref{thm:local_dp_comp}, we can extend the theorems in global DP to the settings of local DP.

\subsection{Disjoint Continual Release}

A DCR against entries has the same properties as that against a database (Definition \ref{def:DCR}). Extending from Theorem \ref{thm:DCR_DP_at_most_k} and \ref{thm:DCR_DP_time_bounded}, we have

\begin{corollary}
    If $\mathbf{q} \in DCR(\epsilon, \delta, \mathbf{t})$ is queried against an entry mutating at-most $k$ times, $\mathbf{q}$ is $2k \text{-} (\epsilon, \delta)$-DP.
\end{corollary}

\begin{corollary}
    If $\mathbf{q} \in DCR(\epsilon, \delta, \mathbf{t})$ is queried against an entry which is $B$-time-bounded in mutations, $\mathbf{q}$ is $2 \text{-} \text{MostSpan}(\mathbf{t}, B) \text{-} (\epsilon, \delta)$-DP.
\end{corollary}

The HDCR against an entry has the same properties as Definition \ref{def:HDCR}. Since a HDCR is the composition of multiple DCRs, given Corallory \ref{cor:nested_k_fold}, we have

\begin{corollary}
    $\text{HDCR}(\epsilon, \delta, h, c, t_s, T, W, M)$ is $2hk \text{-} (\epsilon, \delta)$-DP for an entry mutating at-most $k$ times or $\left[ 2 \sum_{i \in [h-1]}(\frac{1}{c^i} \lceil \frac{B}{W} \rceil + 1) \right] \text{-} (\epsilon, \delta)$-DP for an entry with $B$-time-bounded mutations.
\end{corollary}

\subsection{Sliding-window Continual Release}

A SWCR against entries has the same properties as that against a database (Definition \ref{def:SWCR}). Extending from Theorem \ref{thm:SWCR_DP_at_most_k} and \ref{thm:SWCR_DP_time_bounded}, we have

\begin{corollary}
    If $\mathbf{q} \in \text{SWCR}( \epsilon, \delta, P, W, t_1)$ is queried against a dynamic database whose entries are mutated at most $k$ times, $\mathbf{q}$ is $2 k \text{-} \lceil \frac{W}{P} \rceil \text{-} (\epsilon, \delta)$-DP.
\end{corollary}

\begin{corollary}
    If $\mathbf{q} \in \text{SWCR}( \epsilon, \delta, P, W, t_1)$ is queried against an entry which is $B$-time-bounded in mutations, $\mathbf{q}$ is $2 \lceil \frac{B + W}{P} \rceil \text{-} (\epsilon, \delta)$-DP.
\end{corollary}

\section{Application: Continual Release of Randomized Responses}

In this section, we will outline how to leverage the privacy guarantee of DCR to design a continual release of randomized responses that preserves local differential privacy for dynamic entries. The continual release process will be used to estimate the histogram of the true answers collected from entries over time. 

\subsection{Randomized Response}

Randomized Response Technique \cite{chaudhuri2020randomized} is a common approach to establish local differential privacy of a query \cite{dwork2014privacy}. Suppose a survey asks a question $q$ to an entry $x$, and $r = q(x)$ is the true answer from $x$. However, the entry is reluctant to disclose the true answer, and instead follows a rule $P$ to export a randomized answer (i.e. response) $r' \in \mathbb{R}$, where $\mathbb{R}$ are all the possible answers. Let $|\mathbb{R}| = z$. Let $r_i$ be the $i$-th response in $\mathbb{R}$. The rule $P$ can be represented by a matrix of probability:

\begin{equation}
    P =
    \begin{bmatrix}
        p_{11} & p_{12} & ... & p_{1j} & ... & p_{1z} \\
        p_{21} & ... & ... & ... & ... & ... \\
        ... & ... & ... & ... & ... & ... \\
        p_{i1} & ... & ... & p_{ij} & ... & p_{iz} \\
        ... & ... & ... & ... & ... & ... \\
        p_{z1} & ... & ... & p_{zj} & ... & p_{zz}
    \end{bmatrix}
    \label{eq:P_matrix}
\end{equation}

\noindent where $p_{ij}$ represents the probability of exporting a response $r_i$ given the true answer $r_j$. Also $\sum_{i \in \mathbb{R}} p_{ij} = 1$ for any $j$.

\begin{theorem}
    $P$ is local-$(\epsilon, \delta)$-DP iff for any pair of true answers $r_a$ and $r_b$, and any $S \subseteq \mathbb{R}$, the following inequality is is satisfied:

\begin{equation}
    \sum_{r_i \in S} p_{ia} = e^{\epsilon} (\sum_{r_i \in S} p_{ib}) + \theta .
\end{equation}
\end{theorem}

\noindent The theorem is derived from substituting Eq. \eqref{eq:P_matrix} into the definition of local DP.

\cite{wang2016randomized_response} proves that a randomized response satisfying local $(\epsilon, 0)$-DP can adopt the following rule $P^*$ to optimally utilize the privacy budget:

\begin{equation}
    p_{ij}^* = 
     \begin{cases}
        \frac{e^{\epsilon}}{z - 1 + e^{\epsilon}}, & \text{if } i = j \\
        \frac{1}{z - 1 + e^{\epsilon}}, & \text{if } i \ne j \\
    \end{cases}.
    \label{eq:P_star}
\end{equation}

A randomized response could also be represented using linear algebra. Represent an answer $r_j$ as an $|\mathbb{R}|$-dimensional array $\mathbf{v}_j \in \mathbb{N}^{|\mathbb{R}|}$, where the $j$-th element is one and other elements are zero. Let $\mathbf{U}_j$ be a random variable representing the output of the randomized response $P$ given $r_j$ as input. We have

\begin{equation}
    E[\mathbf{U}_j] = P \mathbf{v}_j.
    \label{eq:U_j_eq_Pv_j}
\end{equation}

\noindent where $E[\mathbf{X}] = (E[X_1], E[X_2], \dots)$. The proof is at Appendix \ref{sec:proof_of_U_j_eq_Pv_j}. 

A survey is interested in the histogram of the true answers from a group of entries, which is denoted as  $\mathbf{v} = (v_1, v_2, ..., v_z)^T \in \mathbb{N}^{|\mathbb{R}|}$, where $v_i$ is the number of true answers $r_i$ from the group. The group's answers are processed using the randomized response $P$, resulting in a histogram of randomized responses represented by a random variable $\mathbf{U} \in \mathbb{N}^{|\mathbb{R}|}$. Recall $\mathbf{U}_j$ is the random variable of the randomized output given $r_j$, and the number of true answer $r_j$ from the group is $v_j$, so we have 

\begin{equation}
    \mathbf{U} = \sum_{j \in [|\mathbb{R}|]^{+}} v_j \mathbf{U}_j.
\end{equation}

\noindent Given the Eq. \eqref{eq:U_j_eq_Pv_j}, we have

\begin{equation}
    E[\mathbf{U}] = P \mathbf{v}.
\end{equation}

\noindent Therefore, we can use $\mathbf{U}$ to estimate the histogram of true answers:

\begin{equation}
    \hat{\mathbf{v}} = P^{-1} \mathbf{U},
\end{equation}

\noindent where $\hat{\mathbf{v}}$ is the estimator of the histogram of the true answers from the entries. 

\begin{theorem}
\label{thm:expectation_of_matrix_transformation}
    Suppose $X$ and $Y$ are two random variables of $n$-dimensional arrays. $P$ is a $n \times n$ matrix with non-zero determinant. If $X = PY$, then
    \begin{equation}
        E[X] = P \: E[Y].
    \end{equation}
\end{theorem}

\noindent The theorem is derived from the linearity of the matrix transformation \cite{meyer2000matrix}. Due to space limitations, we will not provide a detailed proof here.

Given Theorem \ref{thm:expectation_of_matrix_transformation}, we obtain

\begin{equation}
    E[\hat{\mathbf{v}}] = P^{-1} E[\mathbf{U}] = P^{-1} P \mathbf{v},
\end{equation}

\noindent so $\hat{\mathbf{v}}$ is an unbiased estimator of $\mathbf{v}$.

\subsection{Continually Estimate the True Histogram}

\begin{figure}[htp!]
\centerline{\includegraphics[width=1.0\columnwidth]{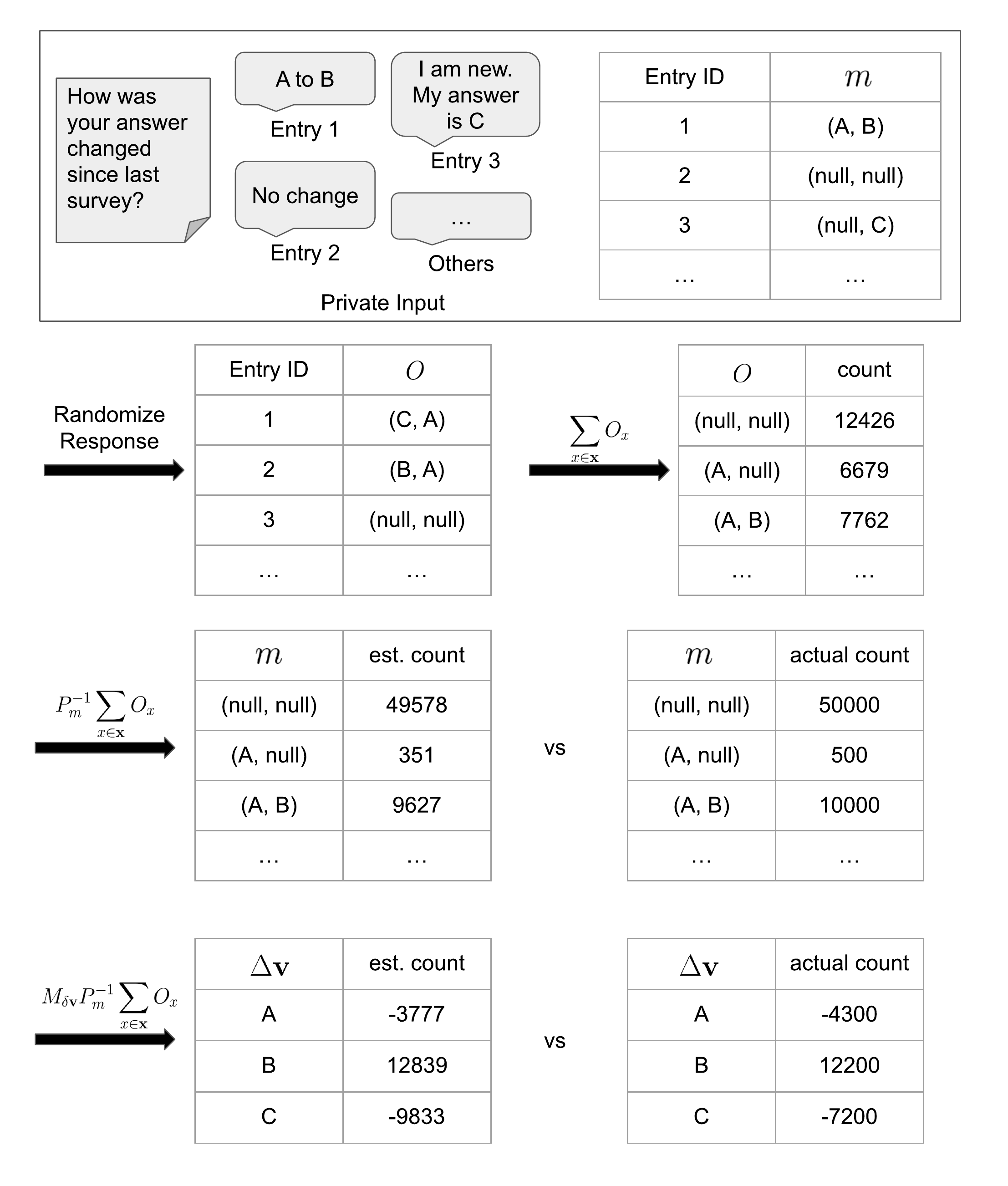}}
\caption{Example of a query of a Randomized Response Continual Release.}
\label{fig: randomized_response}
\end{figure}

A survey may be interested in the evolution of the histogram of the true answers from a group of entries $\mathbf{x}$. Reuse the symbol above: $\mathbf{v}_t$ is the histogram of true answers at time $t$. Let $\Delta \mathbf{v} = \mathbf{v}_{t_2} - \mathbf{v}_{t_1}$ represent the element-wise change of $\mathbf{v}$ from $t_1$ to $t_2$. Obviously, if we can acquire $\Delta \mathbf{v}$ continually, then we can derive the $\mathbf{v}$ continually.

We will show how to derive $\Delta \mathbf{v}$ from the mutations of the true answers of the entries from $t_1$ to $t_2$. Let $m = (m[1], m[2])$ be the mutation of the true answer of an entry from $t_1$ to $t_2$. If there is no change in the answer, $m = (null, null)$. If the answer changes from $r_a$ to $r_b$, then $m = (r_a, r_b)$. If $t_1$ or $t_2$ is out of the lifetime of an entry, $m$ will be $(null, r_b)$ or $(r_a, null)$, respectively. Thus, the space of $m$, denoted by $\mathbb{M}$, is equal to $(\mathbb{R} + \{ \text{null} \})^2$. The contribution of $m$ to $\Delta \mathbf{v}$ can be represented by an $|\mathbb{R}|$-dimensional array $\delta \mathbf{v}$: 

\begin{equation}
    \delta \mathbf{v}[i] = 
    \begin{cases}
      -1, & \text{if } r_i = m[1] \\
      +1, & \text{if } r_i = m[2] \\
      0, & \text{otherwise}
    \end{cases},
    \label{eq:m_to_v}
\end{equation}

\noindent where $\delta \mathbf{v}[i]$ is the $i$-th element of the array $\delta \mathbf{v}$.

Define $m_j$ as the $j$-th element of $\mathbb{M}$. Following the convention of the above section, $m_j$ could also be represented by a $|\mathbb{M}|$-dimensional array, where $j$-th element is one and other elements are zero. Like Eq. \eqref{eq:m_to_v}, $m_j$ is equivalent to $\delta \mathbf{v}_j$, and the conversion from $m$ to $\delta \mathbf{v}$ can be represented by a $|\mathbb{R}| \times |\mathbb{M}|$ matrix

\begin{equation}
    M_{\delta \mathbf{v}} = [\delta \mathbf{v}_1, \delta \mathbf{v}_2, ..., \delta \mathbf{v}_{|\mathbb{M}|}], 
\end{equation}

\noindent so that

\begin{equation}
    \delta \mathbf{v} = M_{\delta \mathbf{v}} m.
\end{equation}

\noindent Let $m_x$ denote the mutation of the true answer of entry $x$, and let $\delta \mathbf{v}_x = M_{\delta \mathbf{v}} m_x$. Given the definition of $\delta \mathbf{v}$, i.e. Eq. \eqref{eq:m_to_v}, and considering all the entries $\mathbf{x}$, we have

\begin{equation}
    \Delta \mathbf{v} = \sum_{x \in \mathbf{x}} \delta \mathbf{v}_x = M_{\delta \mathbf{v}} \sum_{x \in \mathbf{x}}  m_x.
    \label{eq:Delta_v_from_m}
\end{equation}

However, to preserve local DP, $m$ should be processed with a randomized response. Let $P_m$ be the probability matrix of the randomized response whose input is $m$ and output is a random variable $O \in \mathbb{M}$. Both $m$ and $O$ are in form of $|\mathbb{M}|$-dimensional arrays. The value of $P_m$  can be referred to Eq. \eqref{eq:P_star} if it is $\epsilon$-DP. Given Eq. \eqref{eq:U_j_eq_Pv_j}, we have

\begin{equation}
    E[O] = P_m m .
    \label{eq:expectation_O}
\end{equation}

\noindent Let $O_x$ denote a randomized response variable of $m_x$. Define 

\begin{equation}
    \hat{\Delta \mathbf{v}} = M_{\delta \mathbf{v}} P_m^{-1} \sum_{x \in \mathbf{x}} O_x
\end{equation}

\noindent to an estimator of $\Delta \mathbf{v}$. Given Theorem \ref{thm:expectation_of_matrix_transformation}, Eq. \eqref{eq:Delta_v_from_m}, and Eq. \eqref{eq:expectation_O}, 
\begin{multline}
    E[\hat{\Delta \mathbf{v}}] = M_{\delta \mathbf{v}} P_m^{-1} E[\sum_{x \in \mathbf{x}} O_x] =  M_{\delta \mathbf{v}} P_m^{-1} \sum_{x \in \mathbf{x}} P_m m_x \\ = M_{\delta \mathbf{v}} \sum_{x \in \mathbf{x}} m_x = \Delta \mathbf{v},
\end{multline}

\noindent which proves $\hat{\Delta \mathbf{v}}$ is an unbiased estimator of $\Delta \mathbf{v}$. Fig. \ref{fig: randomized_response} shows an example to compute $\hat{\Delta \mathbf{v}}$ from the randomized responses from a group of entries.

$\hat{\Delta \mathbf{v}}$ only depends on the mutations of a group of entries in its timespan. Thus, the continual release of $\hat{\Delta \mathbf{v}}$ satisfies the definition of a Disjoint Continual Release, i.e. $DCR(\epsilon, \delta, \mathbf{t}, \hat{\Delta \mathbf{v}})$. For any $t \in \mathbf{t}$, the sum of $\hat{\Delta \mathbf{v}}$ not later than $t$ yields an unbiased estimator of $\mathbf{v}_t$.

Moreover, $\hat{\Delta \mathbf{v}}$ is a linear query (Section \ref{sec:linear_query}). Thus, it is aggregatable and the continual release of the estimation of $\mathbf{v}_t$  (sum of $\hat{\Delta \mathbf{v}}$) could be derived from a hierarchical disjoint continual release, where each node computes the $\hat{\Delta \mathbf{v}}$ at the timespan of the node. The variance of $\mathbf{v}_t$ is $O(\log t)$ of the variance of $\hat{\Delta \mathbf{v}}$, assuming the start time of the HDCR is zero. The variance of $\hat{\Delta \mathbf{v}}$ is derived in Appendix \ref{sec:variance_of_hat_delta_v}.

\appendices

\section{Proof of Theorem \ref{thm:aggregatable_from_HDCR}}
\label{sec:aggregatable_from_HDCR}

Without loss of generality, we assume $W=1$ and $h = h_{min}$.

Set $T = r - l$. If the start time of $(l, r]$ is aligned with the start time of a node at height $h-1$, then the range can be covered by the ranges of at most $(c-1) \lceil \log_c T \rceil$ nodes as long as $T < c^h$. It is equivalent to representing a number in $c$-base. For example, $(0, 99] = (0, 10] + ... + (80, 90] + (90, 91] + ... + (98, 99]$, and it can be covered by 18 nodes with $c = 10$ and $h = 2$.

Now we only consider the case where $l$ is not the start time of any nodes with height of $h-1$. Split $(l, r]$ to $(l, i] + (i, r]$, where $i$ is the oldest start time of a node with height of $h-1$ in $(l, r]$, i.e. $i = \min_{j \in (l, r]} j \MyMod{c^{h-1}} = 0$. Based on the paragraph above, $(i, r]$ can be aggregated from at most  $(c-1) \lceil \log_c (r - i) \rceil$ nodes.

What about $[l, i)$? Let's first introduce the symmetry of a HDCR: when there is node covering $(i, i + c^j W]$, there is another node covering $(i - c^j W, i]$. Therefore, the number of nodes to cover $[l, i)$ should be identical to the number to cover $[i, 2 i - l)$. Similar to above, $(l, i]$ can be aggregated from at most  $(c-1) \lceil \log_c (i - l) \rceil$ nodes. 

Define $F(l, r)$ to be minimal number of nodes to cover $(l, r]$:

\begin{equation}
    F(l, r) \le (c - 1) \lceil \log_c (i - l) \rceil + (c - 1) \lceil \log_c (r - i) \rceil.
    \label{eq:aggregatable_from_HDCR_step_1}
\end{equation}

\noindent Given $r - i < r - l \le c^h$, we have $\lceil \log_c (r - i) \rceil < h$. Similarly, we have $\lceil \log_c (i - l) \rceil < h$. Then Eq. \eqref{eq:aggregatable_from_HDCR_step_1} becomes

\begin{equation}
    F(l, r) \le 2(c - 1)h . \qed
\end{equation}

\subsection{Proof of Theorem \ref{thm:HDCR_vs_SWCR_at_most_k} and \ref{thm:HDCR_vs_SWCR_time_bounded}}
\label{sec:proof_of_HDCD_vs_SWCR}

Based on Corollary \ref{cor:HDCR_var}, the variance of the HDCR being lower than that of the SWCR is equivalent to

\begin{equation}
    2 (c - 1) h_{min} O(\frac{1}{\epsilon'^2}) < O(\frac{1}{\epsilon^2}).
\end{equation}

\noindent with slight moving terms, and we have 

\begin{equation}
    2 (c - 1) h_{min} O(\epsilon^2) < O(\epsilon'^2).
\end{equation}

For a database whose entries mutate at most $k$ times, when the privacy loss of the HDCR is the same as the SWCR, it requires $\epsilon' = \lceil \frac{W}{P} \rceil \frac{\epsilon}{\lceil \log_c \frac{W}{\Delta T} \rceil} \epsilon$. Also $h_{min} =  \lceil \log_c \frac{W}{\Delta T} \rceil$. Then we derive the Eq. \eqref{eq:HDCR_vs_SWCR_at_most_k}.

For a database being $B$-time-bounded in mutations, when the privacy loss of the HDCR is the same as the SWCR, it requires $\epsilon' = \frac{\lceil \frac{W + B}{P} \rceil}{\frac{c^h - 1}{c^h - c^{h-1}} \lceil \frac{B}{\Delta T} \rceil + h} \epsilon$. Also $h_{min} =  \lceil \log_c \frac{W}{\Delta T} \rceil$. Then, we derive the Eq. \eqref{eq:HDCR_vs_SWCR_time_bounded}.

\section{Proof of Eq. \eqref{eq:U_j_eq_Pv_j}}
\label{sec:proof_of_U_j_eq_Pv_j}

Let $\mathbf{U}_j[i]$ denote the $i$-th element in $\mathbf{U}_j$. $Pr(\mathbf{U}_j[i] = 1) = P_{ij}$, while $Pr(\mathbf{U}_j[i] = 0) = 1 - P_{ij}$. Therefore, $E[\mathbf{U}_j[i]] = P_{ij} = P_i \mathbf{v_j}$, where $P_i = (p_{i1}, ..., p_{iz})$.

Extend this conclusion to all the responses, and we have $E[\mathbf{U}_j] = P \mathbf{v_j}$.

\section{Variance of $\hat{\Delta \mathbf{v}}$}
\label{sec:variance_of_hat_delta_v}

Suppose $X$ is a vector of variables and $A$ is a constant matrix, based on \cite{meyer2000matrix}, we have

\begin{equation}
    Cov(A X) = A Cov(X) A^T,
\end{equation}

\noindent where $Cov(\cdot)$ denotes the covariance matrix of the input. Therefore, $Cov(\hat{\Delta \mathbf{v}})$ can be represented as

\begin{multline}
    Cov(\hat{\Delta \mathbf{v}}) = Cov(M_{\delta \mathbf{v}} P_m^{-1} \sum_{x \in \mathbf{x}} O_x) \\ = M_{\delta \mathbf{v}} P_m^{-1} Cov(\sum_{x \in \mathbf{x}} O_x) P_m^{-1,T} M_{\delta \mathbf{v}}^T.
\end{multline}

\noindent Define $O_{\mathbf{x}} = \sum_{x \in \mathbf{x}} O_x$, and \cite{chaudhuri2020randomized} shows that 

\begin{equation}
    Cov(O_{\mathbf{x}}) = \frac{1}{|\mathbf{x}|} diag(O_{\mathbf{x}}) - O_{\mathbf{x}} O_{\mathbf{x}}^T,
\end{equation}

\noindent where $diag(\cdot)$ denotes the diagonal matrix with the same diagonal elements as the input. Consequently, we have 

\begin{equation}
    Cov(\hat{\Delta \mathbf{v}}) = \frac{1}{|\mathbf{x}|} M_{\delta \mathbf{v}} P_m^{-1}  ( diag(O_{\mathbf{x}}) - O_{\mathbf{x}} O_{\mathbf{x}}^T ) P_m^{-1,T} M_{\delta \mathbf{v}}^T,
\end{equation}

\noindent and $diag(Cov(\hat{\Delta \mathbf{v}}))$ is the variance of $\hat{\Delta \mathbf{v}}$.

% \ifCLASSOPTIONcompsoc
%   % The Computer Society usually uses the plural form
%   \section*{Acknowledgments}
% \else
%   % regular IEEE prefers the singular form
% \fi



\bibliographystyle{IEEEtran}
\bibliography{conference}


% \end{thebibliography}

\end{document}